\documentclass[runningheads]{llncs}
\usepackage{graphicx}
\begin{document}
\title{Audiovisual Affect Assessment and \\Autonomous Automobiles: Applications}
%
%
\author{Bj\"orn W.\ Schuller\inst{1,2,3}\orcidID{0000-0002-6478-8699} \and
Dagmar M.\ Schuller\inst{3}}
\authorrunning{B.\ Schuller and D.\ Schuller}
%
\institute{Imperial College London, London, United Kingdom \and
Chair of Embedded Intelligence for Health Care and Wellbeing, \\University of Augsburg, Germany \and audEERING GmbH, Gilching, Germany \\
\email{schuller@IEEE.org}\\
}
\maketitle              
\begin{abstract}
Emotion and a broader range of affective driver states can be a life decisive factor on the road. While this aspect has been investigated repeatedly, the advent of autonomous automobiles puts a new perspective on the role of computer-based emotion recognition in the car -- the passenger’s one. This includes amongst others the monitoring of wellbeing during the commute such as to adjust the driving style or to adapt the info- and entertainment. This contribution aims to foresee according challenges and provide potential avenues towards affect modelling in a multimodal ``audiovisual plus x'' on the road context. From the technical end, this concerns holistic passenger modelling and reliable diarisation of the individuals in a vehicle. In conclusion, automated affect analysis has just matured to the point of applicability in autonomous vehicles in first selected use-cases, which will be discussed towards the end.

\keywords{Affective Computing  \and Autonomous Vehicles \and Applications.}
\end{abstract}
\section{Introduction}
There is a general acceptance on affect and emotion recognition in vehicles possessing a multitude of highly desirable use-cases \cite{eyben2010emotion,vogel2018emotion}. However, these usually focus on the necessity to monitor the driver in terms of emotions, for example given their safety relevance \cite{hu2013negative}. Accordingly, technical solutions are focusing up to now mainly on the driver in a car \cite{nass2005improving,boril2011utdrive,wollmer2011online,abdic2016driver}. With the advent of autonomous automobiles becoming an every-day reality at scale, however, this focus will naturally shift, as there simply will not be a (human) driver, but rather passengers. We argue that the shift will thereby happen towards the passengers' affect \cite{launonen2021icy}. Unfortunately, there is mainly literature up to this point on passengers' affect in an air transportation context beyond few exceptions \cite{raue2019influence,sini2021passengers}. More insight into passenger's affect in a ground transportation context for autonomous vehicles is therefore urgently needed. 

In the following, we want to share a quick view on how this shift from driver's to passenger's affect in (autonomous) automobiles can lead to novel use-cases. Furthermore, we want to fuel a discussion on what technical requirements will arise from such a shift of focus by highlighting selected dominant implications. For the sake of simplicity, we focus on audio and video as modalities -- however, one can easily think of similar implications for physiology or further modalities.

\section{From Driver's to Passenger's Affect and Emotion Assessment: Applications}
Let us first introduce selected applications of human's affect assessment and an autonomous automobile's reaction to it focusing on the passengers rather than the driver. Note that several of these applications arise from the simple question ``which emotional and social competence does a human driver have that an autonomous automobile is expected to take over''. Secondly, these applications arise from changes of the focus of a passenger not being a driver any longer allowing for other activity during a drive.

\subsection{Driving Style Adaptation}
The authors in \cite{wintersberger2016automated} observe in a driving simulator study that there is no significant influence on the emotion of the passenger seated in the front area of a car by whether a human or an autonomous system steers the vehicle. This may lead to the conclusion that just as a human driver would have to monitor the passengers' wellbeing as influenced by the driving style, an autonomous driving system would have to adapt in similar ways. This could, for example, include aspects such as maximum and average driving speed and acceleration, distance to surrounding vehicles, or radius during turn taking. 

\subsection{Passenger's Affect's Influence on Driving Safety}
With the human driver as controlling element missing who is aware of passengers' affect and its role in the driving safety, it becomes necessary for the autonomous vehicle to do so. This comes, as passengers' affect can be a crucial safety factor. For example, angry infant passengers may pose a risk due to less controlled, but dangerous actions. As an example, children such as seated on rear seats arguing physically during a drive might carry out less controlled actions such as pushing each others or throwing items potentially hitting buttons in the car, damaging the car, or alike. 

\subsection{Passenger's Affect's Influence on Route Planning}
Similar to the point above, a human driver would include the real-time monitoring of her or his passengers' affect to make decisions on the route planning. For example, decreased wellbeing of passengers can lead to the decision to make a stop for resting. In worse cases, a decision may be taken to change the route for a hospital or call for help. In addition, past passengers' affect can be used to make decisions on future route planning. For example, engagement can be exploited to estimate interest in revisiting venues \cite{vitterso2017emotional}. Finally, a passenger's affect can play a decisive role in favouring shorter over longer routes and vice versa \cite{morris2015we}.

However, a car’s preference about the stops, paths, and revisiting venues can of course be co-dependent on the preference of its passengers. And, this passengers' preference might not be completely related to their affective state. Likewise, the decision on these routing parameters should best be made with the passengers, as opposed to for them. In this context, the claims made above may be considered as suggestions, as backing up by literature is partially yet to come.

\subsection{Computer-mediated Communication}
A human driver may also be a mediator in human-to-human conversation, such as when it comes to decide on priorities in route planning. Endowed with emotional competence, future autonomous automobiles may be better equipped for an according task. Furthermore, in case of emergencies such as evacuation of a vehicle in case of an accident, knowledge on a passenger's affect may be of crucial importance. Literature on this issue is mostly available from the air traffic domain, such as \cite{miyoshi2012emergency}. Clearly, however, according considerations of the role of human communication mediation taking affective aspects into account  -- in particular in case of an emergency -- plays a crucial role in ground transportation alike.

\subsection{Entertainment and Information Management}
With the shift from being a driver to being a passenger, there will be more allowance for entertainment and non-driving related activities and information presentation. According choices and suggestions can be based on the passenger's affect, such as affect-based music, movie, or even computer game recommendation \cite{deng2012emotion}, or pre-selection and ordering of information depending on stress level, or other affective states.  

\section{Technical Implications -- Some First Insight}
In a second set of items, we discuss most pressing technical implications arising from the anticipated shift of driver to passengers' affect monitoring in tomorrow's autonomous automobiles. Note, however, that we cannot discuss the technical implications in depth in this communication. 

\subsection{Emotional Target and Context}
With the shift from driver to passenger, the base assumption of most former driver emotion recognition systems is rendered obsolete that the emotion target is mainly the driving situation. In fact, in a situation where passengers are also experiencing entertainment and information presentation during a transportation, or may communicate more with other passengers or remote communication partners, it will become crucial to identify the affect target if exploiting the affect such as in the above use-cases. For example, if the driving style should be adapted depending on a passenger's affect, a system must identify whether the affect is actually induced by the driving style. Such identification of the emotion target has, however, been largely neglected in audiovisual and general affect recognition systems up to now. 

\subsection{Passenger Diarisation and Priorisation}
Given the possibility of multiple passengers in a vehicle, this will require their separation in terms of identifying who is speaking when, or re-identifying individuals in a video feed. As opposed to the present situation in automobiles, where the position in the car defines the role of an individual (driver, non-driver), this assignment of role may vanish in autonomous driving. Further, increasing safety in the future may allow for partial or complete free movement of passengers in tomorrow's autonomous vehicles. Accordingly, the challenge of diarisation of passengers will become significantly more challenging. While passenger diarisation may seem like a trivial problem from today's perspective, especially for video, this may change in the future with passengers moving around more in a car while it is moving, rather than passengers only switching between driving sessions as is mostly the case in today's cars. Re-identification (on video and audio) is also a very mature area of research in computer science, rendering this requirement a more trivial case applying existing techniques. Similarly, detection of multiple faces in the vehicle is addressed to a reasonable degree through existing computer vision algorithms. Lighting and presence of obstacles on the other side is still more of a challenge than distinguishing different faces and can be more challenging with moving passengers. 

At the same time, a prioritisation strategy will be needed for how to cope with affect and emotion as input for decision-making from several individuals in an automobile without a clearly assigned driver. For example, a vehicle may choose to to adapt the driving style to the person feeling the least well. 

\subsection{Passenger Source Separation, Occlusions, and Group Affect}
Related to the above, should there be overlap such as speech overtalk, the different sources will need to be separated in order to be able to assess each passenger's affective state. Similarly, if occlusions in the video feed arise due to potentially increased movement during transportation, technical solutions will be needed to cope with these. The automatic recognition of emotion and affect in such a situation is, however, largely being a white-spot in the literature to-date. Depending on the situation and strategy, the assessment of (passenger) group affect, rather than for each individual may additionally be of interest.

\subsection{Reinforcement Learning}
Affective state recognition has by and large been accomplished by supervised learning from human-labelled data or in a semi-supervised manner. In the above named use-cases, however, there exists an affective feedback-loop allowing for future passengers'  affect-aware autonomous automobiles to learn in a reinforced manner. This enables to learn on a day-to-day basis from thousands of passengers potentially overcoming the field's ever present and dominant bottleneck of sparse learning data. This could include online learning, i.\,e., machine learning in a situation where additional data becomes available during execution of inference with the learnt model. While in the online learning context, many kinds of machine learning technique can be used,  reinforcement learning would be the one where explicit human labelling would not be needed such as by a car asking every now and then its passengers how they feel. Rather, it would learn from their reactions to better assess their affect. For example, assuming a certain affective state such as fear and changing the driving style to slower speed could be followed by a presumably dissatisfied passenger's affect. This could lead a car to revisit the fear assessment made in the first place for future reference. Other than in unsupervised learning, the interaction context would thus be used to better get to know the passengers from an affective perspective without having to ask them explicitly about their emotions and affects. Thus, reinforcement learning can be considered semi-supervised or weakly supervised, but it may reduce active querying of passengers for their affect in order to get to know them better in a potentially more convenient manner.

As a more concrete example, an according vehicle can adapt its driving style or music recommendations based on estimated passengers' affective state. It can then observe the change in passengers' affective state and as a reward function aim at long-term maximum passenger wellbeing. Likewise, a system can improve from interaction with the passengers both on optimising the recognition of passengers' affect and the optimal response patterns.

Note that in the very limited literature on reinforcement learning for improved affect modelling, it could already be shown to be highly effective \cite{Rudovic19-MALa}. 

\section{Discussion and Future Work}

Here, we provided a short listing of potential use-cases, which can neither be complete nor fully comprehensive at all times. All of the use-cases aim to use some degrees of affect to improve users' experience of autonomous vehicles. However, all of the ideas next need to be discussed in depth. In the process of researching these use-case, it will be crucial to further align these (e.\,g., revisiting venues, route planning discussions, entertainment, etc.) with the main goal of an autonomous vehicle: to take passengers to a destination in a safe way. For example, one error that today's human drivers could make is to rely potentially too much on the affective state of the passengers, e.\,g., as discussed above, to change speed and acceleration based on passengers' preferences. Therefore, a future concern will be to adapting driving style to passengers' affective states in a well-balanced manner: Obviously, safety will be first, and the car's driving behaviour should change according to passengers' affective states, only if safety is not affected.

The sketched use-cases are next to be connected with each other leading to an overarching structure. To this end,  a model or framework, even a simple theoretical one, would help to link all the possible arising research areas. Having such a framework will then be useful when designing road maps for the research on the future needs of affect assessment in autonomous automobiles. It will also be helpful in  identifying potentially missing further important use-cases.

More thought will further be needed on the technical issues, which in this contribution are sketched in an abstract manner at an early stage. Oncoming work will need to focus on actual technical models or proposals of such models. Similar to a model for the use-cases, an early first step will be to aim at a broader theoretical framework also for the technical aspects.

Furthermore, the passenger's affect depends also on their interactions with the vehicle, and importantly, on how much they trust the vehicle. Trust will likewise be a very important moderator that is not purely affective, but will influence passengers' affect and how passengers react to the vehicle's actions. Likewise, questions will arise such as should an autonomous automobile drive more slowly and safely to increase trust (and also, reduce the anxiety of the passenger, even though the drive may be taking longer)? Accordingly, the research into affect in autonomous automobiles will also need to include a model of other psychological variables such as trust.
 
Future work will then have to face the changing requirements of affect assessment's application in automobiles ultimately paving the way for increased passenger comfort and wellbeing during autonomous transportation. As automobiles are gradually converging towards full autonomy, and passengers might always want to be able to take part in the driving part at times or to a certain extent, it will further be interesting to also add considerations on affect and its role in semi-autonomous driving.  

\section{Conclusion}

This paper highlighted some interesting perspectives for discussion and future research on automatic affect assessment and its application in autonomous automobiles of the oncoming generations.

We argued that with the advent of autonomous automobiles at scale, the focus on driver affect and emotion will shift to passenger's affect and emotion for applications in emotionally intelligent automobiles. A number of exemplary applications and technical implications were discussed. However, only a small selection of most pressing such could be featured herein. Also, some of the claims made above will need further support such as by oncoming user studies. In the course of gaining more insight into actual requirements and passenger preferences, some of the ideas and use cases above will likely need to be revised. Independent of that, however, and overall, one can expect a major shift in affect recognition and exploitation systems in (autonomous) automobiles of the future. 

\section{Acknowledgement}

The authors acknowledge funding from the European Union's Research \& Innovation Actions (RIA) sustAGE: Smart environments for person-centered sustainable work and well-being (grant no.\ 826506) and WorkingAge: Smart Working environments for all Ages (grant no.\ 210487208).

The authors would like to thank the anonymous reviewers for their valuable contributions which have been reflected in the contribution.

%
%
%
\bibliographystyle{splncs04}
\bibliography{mybib}
\end{document}